\documentclass{webofc}

\usepackage[varg]{txfonts}   
\usepackage{hyperref}
\usepackage{url}
\hypersetup{colorlinks=true,citecolor=blue,urlcolor=blue,linkcolor=blue}
\begin{document}
\title{A weak entanglement approximation for nuclear structure: review and recent developments}
%
%

\author{\textit{Calvin} W. Johnson\inst{1}\fnsep\thanks{\email{cjohnson@sdsu.edu}} }

\institute{Department of Physics, San Diego State University, San Diego, CA, USA}

\abstract{The nuclear shell model is a useful and widely used tool for nuclear structure, 
but it can be hampered by the exponential growth of the basis. Drawing inspiration from 
quantum information theory, one can show that  the proton and neutron components are 
 typically weakly entangled. This has led to the Proton
And Neutron Approximate Shell-model (PANASh). I review the underlying ideas and
present  recent developments. In particular I show how PANASh can accelerate 
beyond-mean-field methods such as the generator coordinate method.}
\maketitle
\section{Introduction}
\label{intro}
There are many models of nuclear structure, but a long-standing one, 
useful for its flexibility and ability to generate excited states, is the interacting 
shell model, also known the the configuration-interaction method~\cite{ca05}.  One expands the wave function in a basis,
\begin{equation}
| \Psi \rangle = \sum_{\alpha} c_\alpha | \alpha \rangle,
\end{equation}
and then finds the stationary states by solving a matrix eigenvalue problem. 

A fundamental question is the choice of basis, $\{ | \alpha \rangle \}$.  
One can choose very simple basis
states, for example Slater determinants, for which
there are fast methods to compute Hamiltonian matrix elements
on-the-fly~\cite{BIGSTICK}, but the basis dimensions needed to reproduce physical features grows exponentially 
with the number of orbitals and the number of particles. 
(Correlated basis states reduce the dimensions, but at a price of much 
more expensive  matrix elements of the Hamiltonian.)
Because the nuclear Hamiltonian is
rotationally invariant,  many nuclear configuration-interaction codes work with
bases with fixed $J_z$ or $M$, called the $M$-scheme. The current largest
$M$-scheme calculations have dimensions of around $10^{10-11}$~\cite{mccoy2024intruder,dao2025exact}. Yet  systems of interest can have
dimensions far beyond this limit. 

Computationally intractable dimensions lead one truncation schemes.
Ideas from quantum information
theory~\cite{johnson2023proton}  inspired a 
  recent
approach~\cite{PhysRevC.110.034305}:  breaking the problem into two pieces,
solving independently, and then combining, leads to an effective and practical
truncation that can extend the reach of the configuration-interaction
shell-model. In Section~\ref{WEA}, I introduce the motivation and
formalism for a ``weak entanglement approximation,'' followed by some sample
results in Section~\ref{results}. This approach has uses beyond the  shell model: applying these ideas to a generator-coordinate calculation 
significantly improves results.

\section{Proton-neutron entanglement in the shell model.}
\label{WEA}

The nuclear shell-model basis states are typically partitioned into proton and neutron components: $ | \alpha \rangle = | a
\rangle_\pi \otimes | i \rangle_\nu$. 
I use indices $a,b$ for the first (proton) components and $i,j$ for the second (neutron) components.
This in turn allows one to exploit ideas taken from
quantum information theory.  The so-called density matrix $\rho_{\alpha,
\beta} = c_\alpha c^*_\beta$ can also be written using these bipartite indices,
$\rho_{a i, b j} = c_{a i} c^*_{b j}$; then one can compute the reduced density
matrix by tracing over one of the partition indices:
\begin{equation}
\rho^\mathrm{red}_{a,b} = \sum_i \rho_{a i, b i} = \sum_i c_{a i} c^*_{b i}. \label{reduced}
\end{equation} 
One can find the eigenvalues of the reduced density matrix, which is nothing
more than singular value decomposition (SVD), also called Schmidt
decomposition; by the SVD theorem  it does not matter over which
partition index we trace.  While the trace of both $\rho$ and
$\rho^\mathrm{red} =1$, the eigenvalues of the former are 0 and 1, while the
eigenvalues $\lambda_r$ of the latter can be on the interval $(0,1)$. The
eigenspectrum can be characterized by the \textit{entanglement entropy},
\begin{equation}
S = - \sum_r \lambda_r \ln \lambda_r.
\end{equation}
$S=0$ means an unentangled system,which can be written as a simple product
wave function. A system with a low $S$, relative to the maximum, is 
``weakly entangled.'' This is not the same as weakly coupled; a system can be
strongly coupled yet weakly entangled, for example, in mean-field calculations. 

Numerical experiments have shown that realistic shell-model wave functions have
low entropy, driven in part by shell structure~\cite{johnson2023proton};
indeed, compared to many other possible partitions of the basis space,
proton-neutron partitioning leads to the lowest
entropy~\cite{perez2023quantum}.  Furthermore, $N \neq Z$ systems have
significantly lower entropy than $N=Z$. This is good news, as heavier nuclides
which are more challenging to model are typically neutron-rich. 

As an example, I computed the proton-neutron entanglement entropy for 
$^{48}$Cr and $^{60}$Cr in the $pf$ valence space, using the $G$-matrix based
$pf$-shell  interaction GXPF1A~\cite{PhysRevC.65.061301}. In this space,
$^{48}$Cr has four valence neutrons while $^{60}$Cr has four valence neutron
holes, meaning they have the same total and component dimensions. 
The $Z=N$ nuclide $^{48}$Cr has an entanglement entropy of 2.84, while $^{60}$Cr 
has an entanglement entropy of 1.84, out of a maximum entropy for these spaces of 8.48.
For more examples see ~\cite{johnson2023proton}.

\begin{figure}[h]
\centering
\includegraphics[width=8cm,clip]{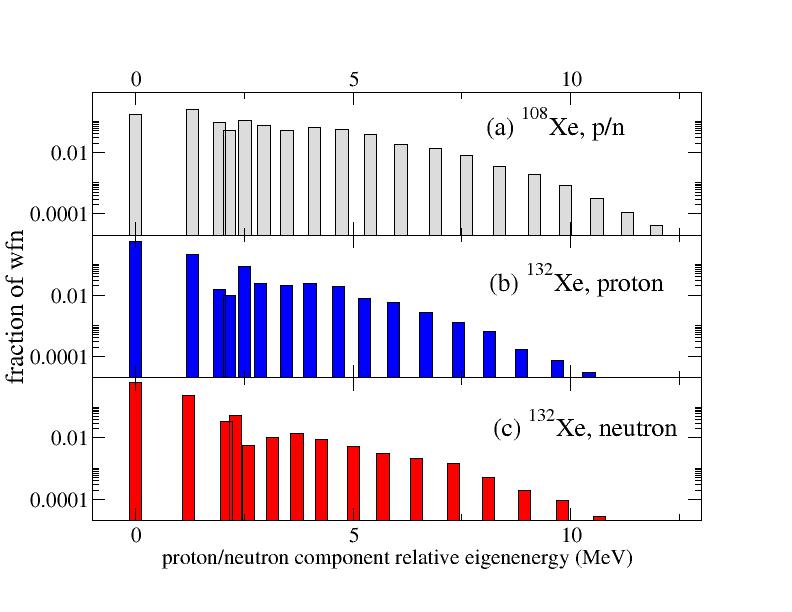}
\caption{Decomposition of configuration-interaction wave functions for select xenon isotopes in the valence space between magic numbers 50 and 82: the fraction of the wave vector projected onto eigenstates of the many-proton/many-neutron components: 
(a) decomposition of $^{108}$Xe; because $Z=N$, the proton and neutron decompositions are identical; (b) proton decomposition of $^{132}$Xe; (c) neutron decomposition of $^{132}$Xe.
 Note the the fast fall-off for 
    $^{132}$Xe, consistent with a lower entanglement entropy. }
\label{fig:xedecomp}       
\end{figure}

To exploit the  weak entanglement between the proton and neutron partitions
(see~\cite{PhysRevC.110.034305} for details), one  expands in a tensor product
basis:
\begin{equation}
| a \, J_a, i   \,  J_i: J \rangle = \left [ | a  \,  J_a \rangle_\pi \otimes | i   \, J_i \rangle_\nu \right ]_J, \label{pnbasis}
\end{equation}
where $| a   \,  J_a \rangle_\pi $ is a many-proton state with angular momentum
$J_a$ and label $a_\pi$, $| i  \,  J_i \rangle_\nu $ is a many-neutron state
with angular momentum $J_i$ and label $i_\nu$, coupled up to some total angular
momentum $J$;
the indexing scheme $a$, $i$ is the same as in Eq.~(\ref{reduced}).
(Parity is suppressed for
clarity.) Working in such a  $J$-scheme (fixed total angular momentum $J$) basis, one expands 
\begin{equation}
| \Psi, J \rangle = \sum_{a,i} c_{a, i} |  a   \,  J_a, i  \,  J_i: J \rangle. \label{pnwfn}
\end{equation}
Using all possible states $a, i$  would recover the full configuration
interaction (FCI) space. 

Rather than taking all possible states, one can truncate  using only a select
set of the proton and neutron components.  This is not a new idea, but unlike
in some previous investigations  which iteratively optimized the
basis~\cite{papenbrock2003factorization,papenbrock2004solution,papenbrock2005density},
we  opt for a ``good enough'' basis.  This is justified by  a
straightforward  investigation.  One divides the shell-model Hamiltonian
into proton, neutron, and proton-neutron sub-Hamiltonians, $\hat{H} = \hat{H}_p
+ \hat{H}_n + \hat{H}_{pn}$ (where $\hat{H}_p$ contains both one-body and
two-body contributions, and same for $H_{n}$; $\hat{H}_{pn}$ is 
only two-body). One can solve the proton and neutron Hamiltonians separately,
\begin{equation}
  \hat{H}_p | \phi_a, J_a \rangle_\pi 
  = E_a | \phi_a, J_a \rangle_\pi, \,\,\,\, \,\,\, \hat{H}_n | 
  \phi_i, J_i \rangle_\nu = E_i | 
  \phi_i, J_i \rangle_\nu; \label{pneigen}
\end{equation}
these proton and neutron eigenstates can be used to construct the basis as in
Eq.~(\ref{pnbasis}).  From the full proton-neutron wave vector,
Eq.~(\ref{pnwfn}),   the fraction associated with each proton (or
neutron)  eigenstate can be found, expressed as a function of the proton-sector
eigenenergy,
 \begin{equation}
f(a) = f(E_a) = \sum_i \left | c_{a,i} \right |^2.
\end{equation}
Even without explicit construction of this choice of basis, one can efficiently
carry out this decomposition using a version of the Lanczos
algorithm~\cite{PhysRevC.91.034313}.  In Fig.~\ref{fig:xedecomp}, I decompose the FCI wave vectors for $^{108}$Xe and
$^{132}$Xe computed in the valence space between magic numbers 50 and 82, that is, the valence space defined by the orbitals $0g_{7/2}$-$2s_{1/2}$-$1d_{3/2,5/2}$-$0h_{11/2}$ valence space using the GCN5082 empirical interaction matrix elements~\cite{Caurier:2007wq,Caurier:2010az}.
$^{108}$Xe has four valence neutrons while $^{132}$Xe has four valence neutron
holes, meaning they have the same total and component dimensions.  Overall one sees an
approximately exponential decrease in the component amplitudes, with a faster
decline associated with the $N > Z$ nuclide, along with a lower entropy. This
behavior is representative of a broader trend.

This exponential decay of component amplitudes leads to a practical methodology. 
One chooses the states $ |  a \, J_a \rangle_\pi$ to be eigenstates of
$\hat{H}_p$, and  the states $| i \, J_i \rangle_\nu$  eigenstates
of $\hat{H}_n$, truncating on the
basis of the energies of the proton and neutron components, which is justified by results such as Fig.~\ref{fig:xedecomp}. (A similar approach was followed by \cite{PhysRevC.92.034320}.) 
This leads to a $J$-scheme code, where
the remaining  key proton-neutron matrix elements coupling the two components
can be computed using one-body
transition density matrices; see ~\cite{PhysRevC.110.034305} for details.  
  The
required inputs (eigenenergies, one-body transition densities) can be produced as a matter of course in an $M$-scheme
code such as {\tt BIGSTICK}~\cite{johnson2018bigstick}. The truncated $J$-scheme dimensions, however, are far smaller, %
though, unlike in the sparse-matrix $M$-scheme calculations of {\tt BIGSTICK}, here the $J$-scheme Hamiltonian matrix is generally fully dense.  The time-to-solution for the PANASh calculation is comparable to or faster than the traditional truncated SM calculation, although 
currently the {\tt PANASh} code is not as fully optimized as {\tt BIGSTICK}. 

\section{Results}
\label{results}

\begin{figure}[h]
\centering
\includegraphics[width=8cm,clip]{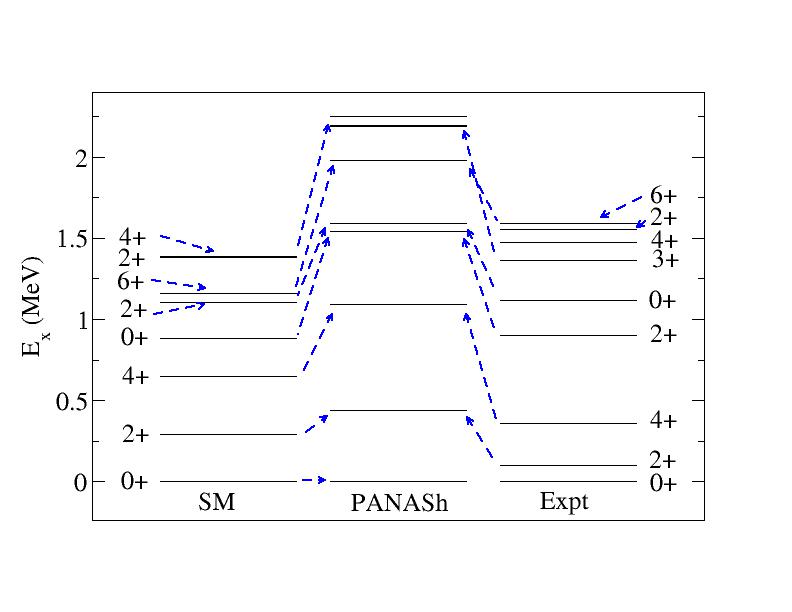}
\caption{Excitation spectrum of $^{130}$Ba in the 50-82 valence space, using the GCN5082 empirical interaction. I compare experiment against a truncated shell-model calculation (SM) with the {\tt BIGSTICK} code, allowing no more than two nucleons to be excited out 
of the $0g_{7/2}$ and 
$1d_{5/2}$ orbitals into the $1d_{3/2}$, $2s_{1/2}$ and $0h_{11/2}$ orbitals. The PANASh calculation uses  component states up to 6 MeV in excitation, or 500 proton component states and 1000 neutron component states. Not shown are the ground 
state energies: -346.8 MeV for PANASh, and -344.3 MeV for SM.}
\label{fig:ba130ex}       
\end{figure}

I apply PANASh to the computation of $^{130}$Ba in the 50-82 valence space, using the GCN5082 interaction.  The FCI $M$-scheme dimension is 220 billion, far beyond what is currently tractable. Fig.~\ref{fig:ba130ex}, I compare three excitation spectra: the experimental excitation spectra,  a PANASh calculation using
500 proton components and 1000 neutron components, which correspond to approximately 6 MeV in excitation in their respective spaces, and a truncated ordinary shell-model calculation, labeled `SM.' For the truncated SM calculation, I allowed at most 2 nucleons to be excited from the $0g_{7/2}$ and 
$1d_{5/2}$ orbitals into the $1d_{3/2}$, $2s_{1/2}$ and $0h_{11/2}$ orbitals, with an $M$-scheme dimension of 760 million, equivalent to a $J$-scheme dimension for the $0^+$ states of about 5.4 million.  In the PANASH calculation, the $0^+$ $J$-scheme dimension is only 24,793, but 
the PANASh ground state energy is -346.8 MeV,  2.6 MeV below the truncated SM ground state energy of -344.3 MeV. Thus, the PANASh calculation is clearly building in important correlations into the ground state, even though the excitation energy of the $2_1^+$ is too high.  
I speculate that the PANASh calculation builds in pairing correlations better than the truncated SM, but may miss out on quadrupole-deformed correlations. By using a deformed mean-field background when generating the proton/neutron component states, future calculations may be able to improve further.

\begin{figure*}
\centering
\vspace*{1cm}       
\begin{tabular}{cc}
\includegraphics[width=6.5cm,clip]{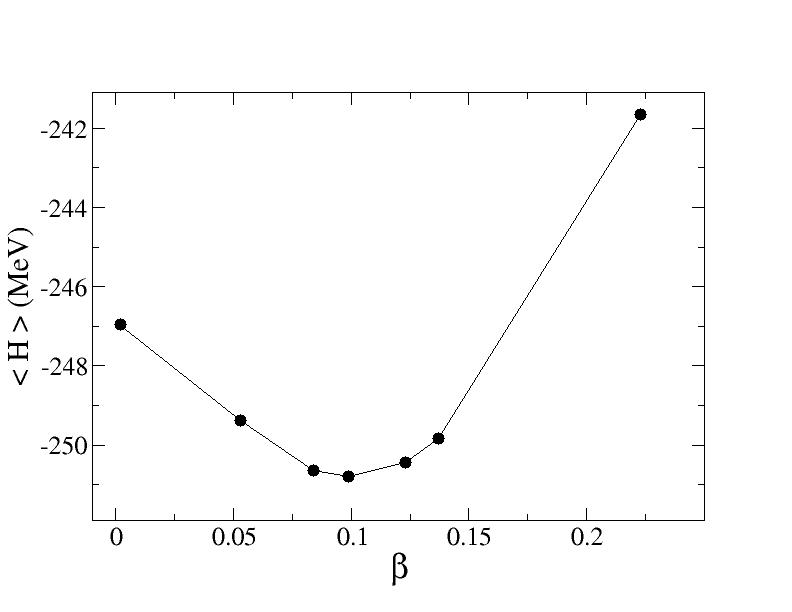}
&
\includegraphics[width=6.5cm,clip]{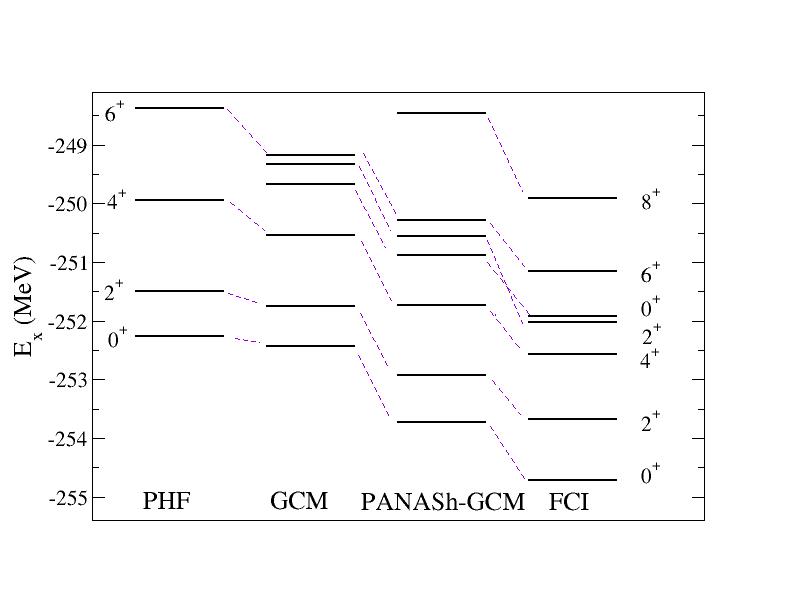}
\end{tabular}
\caption{Improvement of generator-coordinate-method-like methods using PANASh techniques, applied 
to $^{60}$Zn in the $1p0f$ valence space using the GX1A interaction.
Left-hand plot: energies of reference Slater determinants as a function of deformation $\beta$. 
Right-hand plot: Sequence of approximations to low-lying spectra. PHF = angular-momentum projected 
Hartree-Fock using lowest-energy reference state. GCM = diagonalization in subspace defined by projecting 
all seven reference states from left-hand plot.
PANASh-GCM = separately project and diagonalize proton and neutron Slater determinants taken from 
reference states, then recoupled using {\tt PANASh} code.
FCI = full configuration-interaction shell model (full $1f0p$ space) using the {\tt BIGSTICK} code.}
\label{fig:phfpanash}       
\end{figure*}

Finally, one can adapt this formalism to other approaches such as beyond-mean-field methods.  I illustrate this in Fig.~\ref{fig:phfpanash} for the generator coordinate method (GCM)~\cite{ring2004nuclear}, which is applied to 
$^{60}$Zn in the $pf$ valence space using the GXPF1A interaction~\cite{PhysRevC.65.061301,honma2005shell}, which has an $M$-scheme full configuration interaction (FCI) dimension of slightly more than two billion. 
Angular-momentum projected Hartree-Fock (PHF), which projects states of good angular momentum from a single reference Slater determinant state~\cite{ring2004nuclear,lauber2021benchmarking}, provides a mediocre approximation to the excitation spectrum, looking more  a rotational  than 
the actual vibrational-like spectrum. GCM uses additional reference states, generated by constrained Hartree-Fock calculations, in this case by minimizing $\hat{H} + \lambda \vec{Q}\cdot \vec{Q}$, 
where $\vec{Q}$ is the Elliot quadrupole operator.  
The left-hand side of Fig.~\ref{fig:phfpanash} shows the resulting energy landscape as a function of the Bohr deformation parameter $\beta$.  Using seven reference states, including the original minimum Hartree-Fock state, the resulting GCM spectrum is only a little 
better than PHF. One could add more reference states, but the work increases like (number of reference states)$^2$. 

Instead, I adapted the PANASh approach to GCM. Each reference Slater determinant is itself a simple tensor product of a proton and a neutron Slater determinant. From the reference states,  I collected and projected the proton Slater determinants, 
computed the overlaps and Hamiltonian matrix elements, and found the proton eigenstates by a generalized eigenvalue problem, and did the same for the neutron states. After extracting the proton and the neutron 
one-body density matrices, I recoupled the components using the {\tt PANASh} code. The resulting spectrum, labeled as PANASh-GCM in  Fig.~\ref{fig:phfpanash}, is lower in energy and agrees better with the 
numerically exact FCI result, even though I started with the same set of reference states as for the GCM calculation. While a more systematic study is needed, this looks to be a promising way to accelerate GCM-like calculations; the main price to pay is the need for one-body density matrices between the projected eigenstates in the proton and neutron subspaces.

\section{Summary}

One can truncate the nuclear shell model by solving independently the many-proton and many-neutron systems, and then coupling together the low-lying states from each subsystem. This approach is justified by evidence that the proton and neutron components are weakly entangled. 
One gets a better estimate of the ground state energy, and a reasonable approximation to the excitation spectrum, in much smaller spaces than in standard truncations of the shell model. 
 One can also adapt this approach to other methods, such as the generator coordinate method, gaining significant improvements for very little  additional cost.
 
 Near future work will include further optimization of the {\tt PANASh} code. Some preliminary work, not shown here, suggests the basis generation can be improved by including a mean-field from the conjugate component, i.e., generate the proton basis in the presence of a mean-field generated by the neutrons, and vice versa.

\section{Acknowledgements}

 This material is
 based upon work supported by the U.S. Department of Energy, Office of
Science, Office of Nuclear Physics, under Award Number DE-FG02-03ER41272.  This research was enabled by computational resources supported by a gift to
SDSU from John Oldham.

%
\bibliography{johnsonmaster} 
\end{document}